\documentclass[twocolumn,11pt]{article}

\usepackage{xcolor}
\usepackage{lscape}

\usepackage{rotating,booktabs}

\let\subparagraph\relax

\usepackage[top=2cm, bottom=2cm, left=1.8cm, right=1.8cm]{geometry}
\usepackage{times}

\usepackage[small,bf]{caption}
\usepackage{stfloats}
\usepackage{paralist}
\usepackage{enumitem}
\usepackage{xspace}
\usepackage{url}
\usepackage{graphicx}
\usepackage{latexsym}
\usepackage{amssymb}
\usepackage{amsfonts}
\usepackage{psfrag}
\usepackage{wrapfig}
\usepackage{comment}
\usepackage{listings}
\usepackage{paralist}
\usepackage{epstopdf}
\usepackage{tablefootnote,footnote}
\usepackage{flushend}
\usepackage{indentfirst}
\usepackage{verbatim}

\usepackage{cite}
\usepackage{color, colortbl}
\definecolor{darkgreen}{RGB}{0,50,25}
\definecolor{vlightgray}{RGB}{240,240,240}
\usepackage[bookmarks=false, colorlinks=true, plainpages = false,  linkcolor = darkgreen,   citecolor = darkgreen, urlcolor = darkgreen, filecolor = black]{hyperref}

\usepackage[labelformat=simple]{subcaption}

\makeatletter
\input{t1lmtt.fd}
\@namedef{T1+lmtt}{}
\makeatother

\usepackage[marginal,hang]{footmisc}

\newcommand{\descr}[1]{\medskip\noindent\textbf{#1}}

\newcommand{\descrem}[1]{\vspace{0.05cm}\noindent\textbf{\em #1}}

\makeatletter
\def\url@leostyle{%
  \@ifundefined{selectfont}{\def\UrlFont{}}%
  {\def\UrlFont{}}%
}
\makeatother	
\usepackage[hyphenbreaks]{breakurl}

\begin{document}
\pagenumbering{arabic}
\thispagestyle{plain}

\title{\bf Experimental Analysis of Popular Anonymous, Ephemeral,\\ and End-to-End Encrypted Apps\thanks{A preliminary version of this paper appears in the Proceedings of the 2016 NDSS Workshop on Understanding and Enhancing Online Privacy (UEOP). This is the full version.}}
\author{Lucky Onwuzurike and Emiliano De Cristofaro\\[1ex] University College London\\[0.25ex]{\normalsize\{lucky.onwuzurike.13, e.decristofaro\}@ucl.ac.uk}}

\date{}

\maketitle

\begin{abstract}
As social networking takes to the mobile world, smartphone apps provide users with ever-changing ways to interact with each other. Over the past couple of years, an increasing number of apps have entered the market offering end-to-end encryption, self-destructing messages, or some degree of anonymity. However, little work thus far has examined the properties they offer.
To this end, this paper presents a taxonomy of 18 of these apps: we first look at the features they promise in their appeal to broaden their reach and focus on 8 of the more popular ones. We present a technical evaluation, based on static and dynamic analysis, and identify a number of gaps between the claims and reality of their promises.
\end{abstract}

\section{Introduction}
\label{sec:introduction}

Following Edward Snowden's revelations, privacy and anonymity technologies 
have been increasingly often in the news, with a growing number
of users becoming aware -- loosely speaking -- of 
privacy and encryption notions~\cite{dailydot}. 
Service providers have rolled out, or announced they will,
more privacy-enhancing tools, e.g., support for 
end-to-end encryption~\cite{fortune} and HTTPS by default~\cite{mozilla}.
At the same time, a number of smartphone apps and mobile social networks have entered the market, promising to offer features like anonymity, ephemerality, and/or end-to-end encryption (E2EE).
While it is not that uncommon to stumble upon claims like ``military-grade encryption'' or ``NSA-proof''~\cite{nsa} in the description of these apps, little work thus far has actually analyzed the guarantees they provide. 

This motivates the need for a systematic study of a careful selection of such apps. To this end, we compile a list of 18 apps that offer E2EE, anonymity and/or ephemerality, focusing on 8 popular ones (Confide, Frankly Chat, Secret, Snapchat, Telegram, Whisper, Wickr, and Yik Yak). We review their functionalities and perform an empirical evaluation, based on {\em static} and {\em dynamic} analysis, aimed to compare the claims of the selected apps against results of our analysis.

Highlights of our findings include that ``anonymous'' social network apps Whisper and Yik Yak actually identify users with distinct user IDs that are persistent. Users' previous activities are restored to their device after they uninstall and reinstall them, and information collected by these apps could be used to de-anonymize users. We also find that the ephemeral-messaging app Snapchat  does not always delete messages from its servers -- in fact, previous ``expired'' chat messages are surprisingly included in packets sent to the clients. %
Then, we report that all actions performed by a user on Frankly Chat can be observed from the request URL, which is actually transmitted in the clear. \subsection{Building an Apps Corpus}
\label{features}

We start by building a list of smartphone apps that are categorized as ``anonymous'' on Product Hunt~\cite{product},
and those popular among friends and colleagues. We then look at their descriptions and at {\em similar apps} on the Google Play, and focus on those described as offering end-to-end encryption, anonymity and/or ephemerality, as defined below:%

\begin{itemize}
\item[$\bullet$] {\bf\em Anonymity:} is defined as the property that a subject is not identifiable within a set of subjects, known as the anonymity set~\cite{pfitzmann2010terminology}, e.g., as provided by Tor~\cite{dingledine2004tor} for anonymous communications.
In the context of this paper, the term anonymity will be used to denote that users are anonymous w.r.t. other users of the service or w.r.t. the app service provider. 

\item[$\bullet$] {\bf\em End-to-End Encryption (E2EE):} Data exchanged between two communicating parties is encrypted in a way that only the sender and the intended recipient can decrypt it, so, e.g., eavesdroppers and service providers cannot read or modify messages. %

\item[$\bullet$] {\bf\em Ephemerality:} In cryptography, it denotes the property that encryption keys change with every message or after a certain period. Instead, here ephemerality is used to indicate that messages are not available to recipients from the user interface after a period of time~\cite{nyt}.
For instance, in apps like Snapchat, messages ``disappear'' from the app (but may still be stored at the server) a few seconds after they are read. 
\end{itemize}
{\bf First List.} Our first list contains 18 apps, listed in Table~\ref{table:first}, where we also report their first release date, number of downloads as reported by Google Play Store, the kind(s) of content that can be shared via the apps (e.g., text, videos, files), and whether the apps create persistent social links. Note that our first selection does not include popular apps like WhatsApp, since it attempts, but does not guarantee, to provide E2EE for all users~\cite{whatsapp}.

\begin{table*}[t]
\centering
\resizebox{0.99\textwidth}{!}{
\begin{tabular}{|l|c|c|l|l|c|c|c|c|}
\hline
{\bf App} & {\bf Launched} & {\bf \#Downloads} & {\bf Type} & {\bf Content} &
{\bf Anonymity} & {\bf Ephemerality} & {\bf E2EE} & {\bf Social Links}\\ \hline
20 Day Stranger & 2014 & Unknown& Temporary OSN & Photos and location & Yes & No & No & No\\ \hline 
Armortext & 2012 & 50--100K & Chat (Enterprise) & Text and  files & No & User-defined & Yes & Yes\\ \hline 
BurnerApp & 2012 & 100--500K & Temporary & Call and SMS & N/A & N/A & No & Yes\\ 
&  &  &  numbers &  & &  & & \\
\hline 
{\bf Confide} & 2014 & 100--500K & Chat & Text,  documents, photos & No & {\bf After message} & {\bf Yes} & Yes\\ 
& &  && & & {\bf is read} & &\\\hline 
CoverMe & 2013 & 100--500K & Chat & Text,  voice,  photos,  videos & No & User-defined & Yes & Yes\\ \hline 
Disposable & Unknown & 100--500K & Temporary  & Call and SMS & N/A& N/A & No & Yes\\ 
Number &  &  &  numbers &  & &  & & \\ \hline 

{\bf Frankly Chat} & 2013 & 500K--1M & Chat & Text,  pictures,  videos, & Optional for & {\bf 10s} & No & Yes\\
 & & & & voice & group chat & &  & \\ \hline
{\bf Secret} & 2014 & 5--10M & Anonymous OSN, & Text, photos, & {\bf Yes} & No & No & Yes/No\\ 
 & & & Chat & & & & &\\
\hline 
Seecrypt SC3 & 2014 & 10--50K & Chat & Text, voice, files & No & No & Yes & Yes\\ \hline 
Silent Circle & 2012 & 100--200K & Encrypted Phone & Call, SMS, files & No & User-defined & Yes & Yes\\
\hline 
{\bf Snapchat} & 2011 & 100--500M & Transient OSN & Photos, videos & No & {\bf 1 -- 10s} & No & Yes\\ \hline 
{\bf Telegram} & 2013 & 50--100M & Chat & Text, photos,  audio, & No & {\bf Optional} & {\bf Optional} & Yes\\ 
& & & &  videos, files, location && & & \\ \hline
TextSecure & 2010 & 500K--1M & Chat & Text, files & No & No & Yes & Yes\\ \hline 
TigerText & 2010 & 500K--1M & Chat & Text, files & No & User-defined & Yes & Yes\\ \hline 
Vidme & 2013 & 50--100K & Video Sharing & Videos & Yes & No & No & No\\ \hline 
{\bf Whisper} & 2012 & 1--5M & Anonymous OSN, & Text, photos & {\bf Yes} & No & No & No\\ 
& & &  Chat & & & & &\\\hline 
{\bf Wickr} & 2012 & 100--500K & Chat & Text, files, photos, audio, & No & {\bf User-defined} & {\bf Yes} & Yes\\ 
& & & & videos && & & \\ \hline
{\bf Yik Yak} & 2013 & 1--5M & Local Bulletin & Text & {\bf Yes} & No & No & No\\ \hline
\end{tabular}
}
\vspace{.1cm}
\caption{Our first selection of 18 smartphone apps providing at least one among ephemerality, anonymity, or end-to-end encryption. N/A denotes `Not Applicable'. Apps in bold constitute the focus of our analysis in Sections~\ref{sec:static}-\ref{sec:dynamic}.}
\label{table:first}
\end{table*}

\subsection{Apps Selection}\label{sec:selection}
Among the 18 apps presented in Table~\ref{table:first}, we then select a few popular ones, as discussed below.

\descr{Selection Criteria.} From our corpus, we focus on apps with the most downloads that offer ephemerality, anonymity, E2EE, or, preferably, a combination of them. 
We exclude Silent Circle and TigerText as they require, respectively, paid subscription and a registered company email.
We reduce our selection to 8 apps: Confide, Frankly Chat, Secret, Snapchat, Telegram, Whisper, Wickr, and Yik Yak (bold entries in Table~\ref{table:first}).
Next, we provide an overview of their advertised functionalities, complementing information in the Table~\ref{table:first}.
(Note that descriptions  below are taken either from the Google Play Store or the apps' respective websites.)

\descrem{Confide:} offers end-to-end encryption and ephemerality. It allows users to share text, photos, and documents from their device and integrates with Dropbox and Google Drive. It provides read receipts and notification of screenshot capture attempts. Messages are not displayed on the app until the recipient ``wands'' over them with a finger, so that only a limited portion of the message is revealed at a time. After a portion of the message is read, it is grayed out Screenshots are also disabled on Android. 
Messages that have not been read are kept on the server for a maximum of 30 days.

\descrem{Frankly Chat:} is a chat app allowing users to send ephemeral messages (text, picture, video or audio), anonymous group chats, and un-send messages that the recipient has not opened. 
Messages disappear after 10 seconds but users can ``pin'' their chats disabling ephemerality. Both parties do not need to have the app installed to receive messages: a link is sent to the recipient via email, when clicked, reveals the message. Messages are deleted from the server after 24 hours---whether they are read or not.

\descrem{Secret:} {\em (discontinued May 2015)} lets users post anonymously to other {\em nearby} users. Users can view \textit{secrets} from other locations but can only comment on those from their nearby location. 
Users can chat privately with friends and engage in a group chat with the chat history disappearing after a period of inactivity. 

\descrem{Snapchat:} is an app that allows users send text, photos and videos that are displayed for 1 to 10 seconds (as set by the user) before they ``disappear'', i.e., they are no longer available to their friends. If the recipient takes a screenshot, the sender is notified.
Users can also view \textit{Stories}, i.e., a collection of snaps around the same theme, and a so-called~\textit{Discover}, i.e., accessing snaps from different selected editorials.  

\descrem{Telegram:} is a messaging app that lets users exchange text, photos, videos, and files. It also provides users with an option to engage in a ``secret chat'', which provides E2EE and optional ephemerality. Senders are notified if the recipient takes a screenshot. Account information, along with all messages, media, contacts stored at Telegram servers are deleted after 6 months of login inactivity.

\descrem{Whisper:} is a location-based mobile social network that allows users to anonymously share texts displayed atop images, which are either selected by the users or suggested by the app. Users can view and respond to \textit{whispers} either as a private message or via another \textit{whisper}. 

\descrem{Wickr:} is a chat app supporting text, audio, video, photos, and files, with user-defined ephemerality (maximum 6 days). It also allows users to engage in group chats, shred deleted files securely, and prevents screenshots on Android and claims to anonymize users by removing metadata (such as persistent identifiers or geo-location) from their contents.

\descrem{Yik Yak:} is a local bulletin-board social network allowing nearby users to post \textit{yaks} anonymously. Users clustered within a 10-mile radius are considered local and can post, view, reply to, and up/down vote yaks but can only view \textit{yaks} outside their locality.

\section{Static Analysis}
\label{sec:static}
We now present the results of a static analysis of the 8 apps, aiming to analyze SSL/TLS implementations and look for potential information leakage.

\subsection{Methodology}
We perform static analysis using dex2jar~\cite{dex2jar}, decompiling the .apk files to .jar files, from which we extract the related Java classes using JD-GUI~\cite{jdgui}. We then search for SSL/TLS keywords like \texttt{TrustManager}~\cite{tm}, \texttt{HostnameVerifier}~\cite{hv}, \texttt{SSLSocketFactory}~\cite{sslsf}, and \texttt{HttpsURLConnection} \cite{httpsUrl}.
Then, we inspect the \texttt{TrustManager} and \texttt{HostnameVerifier} interfaces used to accept or reject a server's credentials: the former manages the certificates of all Certificate Authorities (CAs) used in assessing a certificate's validity, while the latter performs hostname verification whenever a URL's hostname does not match the hostname in the certificate.

\subsection{Results}
Several sockets are usually created to transport data to different hostnames in an app, therefore, sockets in an app may have different SSL implementations. We observe different SSL implementations in the 8 apps, and summarize our findings below.

\descr{Non-Customized SSL Implementation.} App developers can choose to use any one of five defined \texttt{HostnameVerifier} subclass for hostname verification, and use \texttt{TrustManager} initialized with a keystore of CA certificates trusted by the Android OS to determine whether a certificate is valid or not, or customize certificate validation by defining their own logic for accepting or rejecting a certificate.
All the 8 apps in our corpus contain some non-customized SSL implementations. Telegram and Yik Yak only use non-customized SSL code, with the former relying on \texttt{BrowserCompatHostnameVerifier} class and building sockets from default \texttt{SSLSocketFactory}.
Confide and Snapchat both use the \texttt{BrowserCompatHostnameVerifier} class, while Wickr has instances of all \texttt{HostnameVerifier} subclasses but uses the \texttt{BrowserCompatHostnameVerifier} class on most of its sockets. Snapchat does not customize its \texttt{TrustManager} either and registers a scheme from the default \texttt{SocketFactory}. Secret uses sockets from the default \texttt{SSLSocketFactory} but employs regex pattern matching for hostname verification.

\descr{Vulnerable TrustManager/HostnameVerifier.} Frankly Chat, Whisper, and Wickr all contain \texttt{TrustManager} and \texttt{HostnameVerifier} that accept all certificates or hostnames. Alas, this makes it possible for an adversary to perform {\em Man-in-The-Middle (MiTM)} attacks and retrieve information sent on the sockets that use the vulnerable \texttt{TrustManager} and/or \texttt{HostnameVerifier}. Vulnerable \texttt{HostnameVerifier} in Frankly Chat returns \texttt{true} without performing any check, while Wickr uses 
the \texttt{AllowAllHostnameVerifier} subclass which is also used in Whisper by Bugsense crash reporter. 

\descr{Certificate Pinning.} Confide, Frankly Chat, and Whisper implement certificate pinning. Confide pins the expected CA certificate which is also accessible from the decompiled apk, whereas, Whisper uses the hash of the pinned certificate appended with the domain name to make certificate validation decisions. For Frankly Chat, a single certificate is expected and its hash is checked for in the received certificate chain. Frankly Chat also initialize another \texttt{TrustManager} with a keystore that loads certificate from file.

\begin{figure*}[t]
\centering
    \begin{subfigure}[t]{0.47\textwidth}
\includegraphics[width = 0.99\columnwidth]{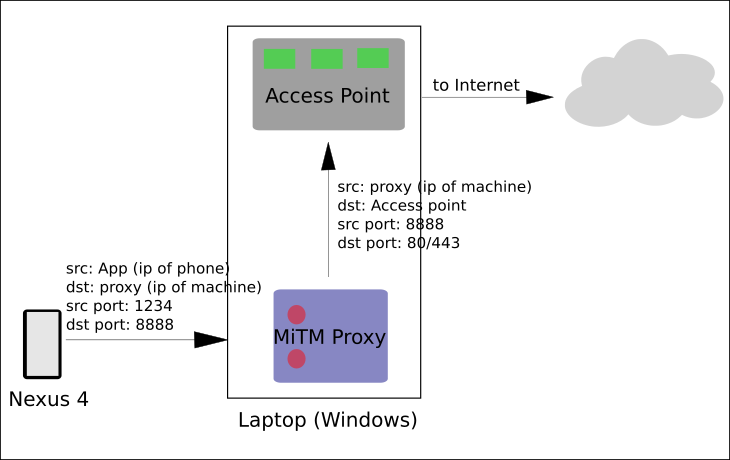} %
\vspace{-0.1cm}
\caption{Proxy Setup Using Fiddler.}
\label{fig:fiddler}
\end{subfigure}
~
    \begin{subfigure}[t]{0.47\textwidth}
\centering
\includegraphics[width = 0.99\columnwidth]{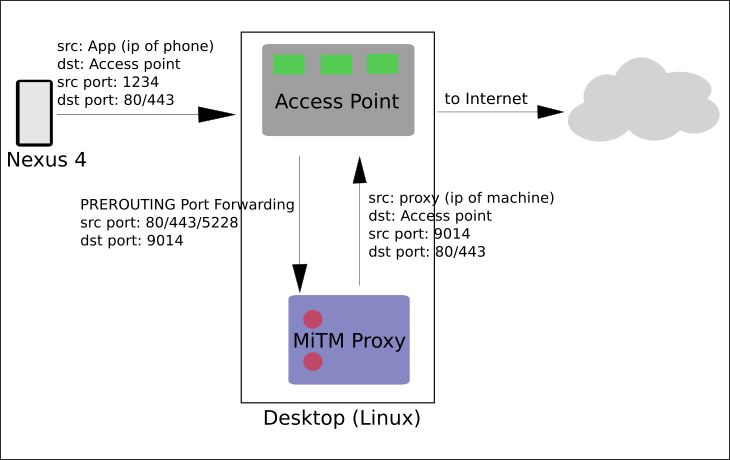} %
\caption{Proxy Setup Using SSLSplit.}
\label{fig:mitmproxy}
\end{subfigure}
\vspace{-0.2cm}
\caption{Dynamic Analysis Setup.}
\end{figure*}

\section{Dynamic Analysis}
\label{sec:dynamic}
Next, we present the results of our dynamic analysis aimed to scrutinize features ``promised'' by the 8 apps in our corpus, as well as to confirm whether the vulnerabilities found statically are also observed dynamically.

\subsection{Experimental Setup}
We conduct our experiments on a LG Nexus 4 running Android 5.1, connected to a Wi-Fi access point under our control. 
(Note that the Wi-Fi network was secured using WPA2 to prevent unauthorized connections and ensure that only intended traffic was captured.)
Our intention is to examine what a random attacker can access from an app advertised as privacy-enhancing, and what can be deduced as regards privacy-enhancing claims. Hence, we assume an adversary that cannot elevate her privilege nor have access to a rooted device. We perform actions including sign-up, login, profile editing, sending/reading messages, while monitoring traffic transmitted and received by the apps. We collect traffic using Wireshark and analyze unencrypted traffic to check for sensitive information transmitted in the clear.
We also rely on HTTP proxies such as Fiddler~\cite{fiddler} and SSLSplit~\cite{mitm} to mount Man-in-The-Middle (MiTM) attacks and decrypt HTTPS traffic. Proxying is supported in two ways:
\begin{enumerate}
\item {\em Regular Proxy:} We install the Fiddler HTTP proxy~\cite{fiddler}
on a Windows 8.1 laptop (which also acts as Wi-Fi access point), listening on port 8888, and manually configure the smartphone to connect to the proxy. Figure~\ref{fig:fiddler} illustrates our proxy setup using Fiddler. We also install Fiddler's CA certificate on the smartphone and laptop to allow  HTTPS traffic decryption.\smallskip
\item {\em Transparent Proxy:} Some Android apps are programmed to ignore proxy settings, so Fiddler does not accept/forward their packets. This happens with Telegram, Wickr (non-CSS/JS), and Frankly Chat (chat packets). Therefore, we set up a transparent proxy as shown in Figure~\ref{fig:mitmproxy} using  SSLsplit MiTM proxy~\cite{mitm} set to listen on port 9014
on a Linux desktop running Fedora 22, which also acts as a Wi-Fi access point.
We use \textit{iptables} to redirect to port 9014 all traffic to ports 80, 443, and 5228 (GCM). As SSLsplit uses a CA certificate to generate leaf certificates for the HTTPS servers each app connects to, we generate and install a CA certificate on the smartphone, and pass it to SSLsplit running on the Linux machine. 
\end{enumerate}

\subsection{Results}
We now present the results of our dynamic analysis, which are also summarized in Table~\ref{table:proxy}. 

\descr{No Proxy.} We start by simply analyzing traffic captured by Wireshark and observe that Secret and Frankly Chat send sensitive information in the clear. Specifically, in Frankly Chat, the Android advertising ID (a unique identifier) is transmitted in the clear, via an HTTP GET request, along with Device Name. The list of actions a user performs on Frankly Chat can also be observed from the request URL. Secret instead leaks Google Maps location requests (and responses) via HTTP GET. 

\descr{Regular Proxy.} Using Fiddler as a MiTM proxy, we notice that Confide and Whisper do not complete connection with their servers due to certificate pinning. Note that Whisper started implementing pinning after an update on April 22, 2015. Prior to that, one could capture Whisper traffic via Fiddler and access location and user ID.
We also notice that Frankly Chat hashes passwords using MD5 without salt, while Snapchat sends usernames and passwords without hashing. Although inconsistently, Snapchat also sends previous ``expired'' chat messages to the other party even though these are not displayed on the UI.\footnote{Note: we have informed Snapchat of this in October 2015.}

Decrypted traffic from Secret, Whisper, and Yik Yak show that these apps associate unique user IDs to each user, respectively, \textit{ClientId}, \textit{wuid}, and \textit{user ID}. We test the persistence of these IDs and find that, even if the apps' cache on the device is cleared through the Android interface, and the apps uninstalled and reinstalled, Whisper and Yik Yak retain the user ID from the uninstalled account and restore all previous \textit{whispers} and \textit{yaks} from the account. On Whisper, we manually delete its wuid and state files (in the /sdcard/whisper directory) before reinstalling the app: this successfully clears all previous \textit{whispers} and a new wuid file is generated. However, it does not completely de-associate the device from the ``old'' account as the ``new'' account would still get notifications of private messages from conversations started by the ``old'' account. On the contrary, clearing Secret's cache unlinks previous messages, even without uninstalling the app. 

Telegram and Wickr ignore the proxy settings, i.e., traffic does not pass through our proxy. Frankly Chat also ignore the proxy when sending chat messages but not for traffic generated by other actions.

\begin{savenotes}
\begin{table}[t]
\centering
\small
\resizebox{0.95\columnwidth}{!}{
\begin{tabular}{|p{0.15\linewidth}|p{0.38\linewidth}|p{0.38\linewidth}|}
\hline
{\bf App} & {\bf Fiddler} & {\bf SSLSplit} \\
\hline
Confide & No connection & No connection \\
\hline
Frankly Chat & TLS traffic is decrypted but packets containing chat messages not routed through proxy & TLS traffic is decrypted but there is no connection to the server when chat is attempted \\
\hline
Secret & All packets decrypted & Not Available (discontinued before we started using the transparent proxy) \\
\hline
Snapchat & All packets decrypted & All packets decrypted \\
\hline
Telegram & Connects but traffic does not pass through proxy & TLS traffic is decrypted but E2EE is enabled \\
\hline
Whisper & No connection & No connection \\
\hline
Wickr & Connects but traffic does not pass through proxy & TLS traffic is decrypted but E2EE is enabled\\
\hline
Yik Yak & All packets decrypted & All packets decrypted\\
\hline
\end{tabular}
}
\caption{Summary of Dynamic Analysis Results.}
\label{table:proxy}
\vspace{-0.2cm}
\end{table}
\end{savenotes}

\descr{Transparent Proxy.} Using SSLSplit, we decrypt SSL-encrypted traffic from Wickr and Telegram. We do not find any sensitive information or chat messages being transmitted as the traffic is indeed encrypted.
Apps for which SSL-encrypted traffic is recovered using Fiddler exhibit the same behavior on the transparent proxy, with Confide and Whisper not connecting due to pinning. We observe that certificate pinning is implemented on the socket used to transmit chat messages on Frankly Chat, as we cannot send chat messages but perform other actions, e.g., editing profiles and adding new friends. We also uninstall the CA certificate from the device to observe whether non-trusted certificate are accepted, and find that none of the apps established an HTTPS connection, which implies the apps do not use TrustMangers accepting any certificate as valid as reported in~\cite{fahl2012eve,onwuzurike2015}.

\section{Discussion}
\label{discussion}

We now discuss the implications of our analysis, in light of the properties promised by the 8 studied apps. %

\descr{Anonymity w.r.t. other users.} Posts on Secret and Yik Yak (resp., \textit{secrets} and \textit{yaks}) are not displayed along with any identifier, thus making users anonymous w.r.t. other users. 
Whereas, on Whisper, due to the presence of a {\em display name} (non-unique identifier shown to other users) and its ``Nearby'' function, communities can be formed as a result of viewing and responding to {\em whispers} from nearby locations. Thus, it may be possible to link \textit{whispers} to a display name, while at the same time querying the distance to the target, as highlighted in~\cite{wang2014whispers}. 

A user who thinks is anonymous is more likely to share sensitive content she might not share on non-anonymous OSN platforms, which makes ``anonymous'' apps potential targets of scammers/blackmailers that can identify users. This motivates us to examine the possibility of creating social links between users, i.e., linking a user and a set of actions. We find that this is not possible on Yik Yak as there are no one-to-one conversations. Also, when the Yik Yak stream is monitored by a non-participating user, user IDs observed are symbolic to the real unique user ID. The symbolic user ID is only associated to one \textit{yak}, hence one cannot use it to link a user as the ID differs across \textit{yaks} by the same user.
Frankly Chat optionally offers $k$-anonymity during a group chat with $k$+1 friends. Due to the social link already present in the group (users chat with friends), psychological differences make it possible to identify who says what. 

\descr{Anonymity w.r.t. service provider.} All apps associate identifiers to its users, which allows them to link each user across multiple sessions. Wickr claims to strip any metadata that could allow them to identify their users, thereby making users anonymous and impossible to track \cite{wickr1}, but we cannot verify this claim since all traffic is encrypted end-to-end. 

We observe different levels of persistence of user IDs in Secret, Whisper, and Yik Yak, as mentioned earlier. Secret stores identifiers on users' device, so an identifier would cease to persist beyond data and cache clearance. Whereas, for Whisper and Yik Yak, we have two hypotheses as to why user IDs survive when the app is uninstalled and later reinstalled: either they store identifiers on their servers and restore them to a device on re-installation, or they create the user IDs from the same device information using a deterministic function. This observation indicates that Whisper and Yik Yak's user IDs are linked to device information, thus making users persistently linkable.
While Whisper and Yik Yak do reveal the information they collect from users in their privacy policy, previous work shows that the overwhelming majority of users do not read (or anyway understand) privacy policies~\cite{internetsociety}. Both apps collect information including device ID, IP address, geo-location, which can be used to track users. This, along with profiles from analytics providers (which both apps embed), can be used to de-anonymize users' age, gender, and other traits with a high degree of accuracy \cite{seneviratne2014predicting}. Finally, note that Whisper's description on Google Play, while including terms like `anonymous profiles' and `anonymous social network', is actually ambiguous as to whether they refer to anonymity w.r.t to Whisper or other users (or both).

\descr{Location Restriction.} Secret and Yik Yak's restriction on feeds a user can see (and interact with) can simply be defeated, e.g., as Android lets users to use mock locations in developer mode. In combination with an app that feeds GPS locations chosen by the user (e.g., {\em Fake GPS}), this allow them to access geo-tagged messages from anywhere.

\descr{Ephemerality.}
Confide, Frankly Chat, Snapchat, Telegram, and Wickr offer message ephemerality with varying time intervals. Confide claims messages disappear after it is read once~\cite{confide} but this is not the case as messages only "disappear" after a user navigates away. This implies the recipient can keep the message for longer as long as they do not navigate away from the opened message. In Frankly Chat, messages ``disappear'' after 10 seconds (even though users can pin messages). Ephemerality on Telegram only applies to ``secret chats'' and the expiration time is defined by the user. Snapchat and Wickr also let users determine how long their message last, with Snapchat defining a range of 1--10s (default 3s). 
On Snapchat, previous chat messages are actually part of the response received from the server, even though they are not displayed on the client's UI. This indicates that read messages are actually not deleted from Snapchat servers immediately, despite what is stated in Snapchat's privacy policy \cite{snapprivacy}. Since Confide and Frankly Chat implement certificate pinning, we cannot examine if responses from the server during chat contain past messages. Also, Telegram and Wickr encrypt data before transmission, thus we cannot make any analysis from intercepted packets.

Of all the apps offering ephemerality, only Confide and Wickr instruct the Android OS to prevent screen capture from a recipient. 
Obviously, however, the recipient can still take a photo with another camera, and video recording would defeat Confide's wand-based approach.
Confide can claim to offer plausible deniability if a photo is taken, as messages are not displayed along with the name of the sender, hence, pictures would not preserve the link between the message and the identity of the sender. 
Frankly Chat, Snapchat, and Telegram only notify the sender that the recipient has taken a screenshot, thus ephemerality claims are only valid assuming the recipient is not willing to violate a social contract between them and the sender.
Also, if messages are not completely wiped from the server, the provider is obviously still subject to subpoena and/or vulnerable to hacking.

\descr{End-to-End Encryption.}
Confide and Wickr claim to employ E2EE by default, using AES-128 and AES-256, respectively. We can confirm E2EE in Wickr but not in Confide, since certificate pinning prevents interception of traffic. Also, Telegram offers E2EE for ``secret chat'' using AES-256 and client-server encryption (i.e. only the server and both clients can decrypt traffic) which also prevents MiTM attacks for non-secret chats. In both secret and non-secret chat, Telegram uses a proprietary protocol, MTProto, and transmit traffic over SSL although its webpage states otherwise.\footnote{\url{https://core.telegram.org/mtproto\#http-transport}} Telegram and Wickr's implementations also claim to support perfect forward secrecy \cite{telegrampfs,wickr1}.

Finally, note that recent criticism of Telegram's security in the press\footnote{\url{http://preview.tinyurl.com/ntahv65}} do not affect the claims of Telegram that we choose to analyze, i.e., E2EE and ephemerality in ``secret chats.''

\section{Related Work}
\label{sec:related}
This section reviews related work, specifically, (i) measurement studies of chat apps and location-based social networks, (ii) apps vulnerabilities, and (iii) investigations of users' behavior.

\descr{Measurement-based studies.} Wang et al.~\cite{wang2014whispers} analyze user interaction in Whisper, motivated by the absence of persistent social links, content moderation, and user engagement. They also highlight a vulnerability that allows an attacker to detect a user's location by attaching a script to a {\em whisper} querying Whisper's DB. 
Correa et al.~\cite{mpi-icwsm} define the concept of {\em anonymity sensitivity} for social media posts and measure it across non-anonymous (e.g., Twitter) and anonymous (e.g., Whisper) services, aiming to study linguistic differences between anonymous and non-anonymous social media sites as well as to analyze content posted on anonymous social media and the extent user demographics affect perception and measurements of sensitivity.
Peddinti et al.~\cite{peddinti2014cloak} analyze users' anonymity choices during their activity on Quora, identifying categories of questions for which users are more likely to seek anonymity. They also 
perform an analysis of Twitter to study the prevalence and behavior of so-called ``anonymous'' and ``identifiable'' users, as classified by Amazon Mechanical Turk workers, and find a correlation between content sensitivity and a user's choice to be anonymous. %
Stuzman et al.~\cite{stutzman2013silent} observe a significant growth in anonymity-seeking behavior on online social media in 2013, while Roesner et al.~\cite{roesner2014sex} analyze why people use Snapchat: they survey 127 adults and find that privacy is not the major driver of adoption, but the {\em ``fun''} of self-destructing messages.

\descr{Flaws.}
Prior work has also looked at related apps' security flaws:
in late 2013, researchers from Gibson Security discovered a flaw in Snapchat's API that allows an adversary to reconstruct Snapchat's user base (including names, aliases, phone numbers) within one day and mass creation of bogus accounts~\cite{zdnet}. Zimmerman~\cite{wickrftt} highlights the issue of linkability of anonymous identifiers in Wickr. Recently, Unger et al.~\cite{sok} systematize security and usability of chat and call apps providing end-to-end encryption. Also, prior work~\cite{fahl2012eve, sounthiraraj2014smv, onwuzurike2015} has studied libraries, interfaces, classes, and methods used by apps to make security decisions, specifically, w.r.t. vulnerabilities in sockets used to transmit user data.

\descr{User Behavior.} Pielot and Oliver~\cite{chi2014} study the motivations behind the use of Snapchat by teenagers. They create two personas and, by engaging with other users, they find that teens use Snapchat as they are excited by the ephemerality, see fewer risks, and non-commitment to persistent messengers.
Roesner et al.~\cite{roesner2014sex} analyze why people use Snapchat: they survey 127 adults and find that security and privacy are not the major drivers of adoption, but rather the {\em ``fun''} of self-destructing messages. 
Hosseinmardi et al.~\cite{hosseinmardi2014towards} look at cyberbullying on a semi-anonymous network, i.e., last.fm, while Stuzman et al.~\cite{stutzman2013silent} observe a significant growth in anonymity-seeking behavior on online social media in 2013.  Shein~\cite{shein2013ephemeral} interview a few experts and commented on the rise of apps for ``ephemeral data'' (e.g., Snapchat, Gryphn, Wickr), pointing out that users do not use ephemeral messaging because they have something to hide, rather, because they do not want to add digital artifacts to their digital ``detritus.''

\descr{Privacy Perceptions.}
Liu et al.~\cite{liu2011analyzing} measured the discrepancy between desired and actual privacy settings of Facebook users, with a user study involving 200 participants. Authors found that perception matched reality only 37\% of the time, and that default settings were used for 36\% of the profiles. 
Ayalon and Toch~\cite{ayalon2013retrospective}  investigated the relationship between information sharing, information aging, and privacy. They conducted a survey of 193 Facebook users and posited that relevance, willingness to share/alter posts decreases with time. They also found that users are more willing to share recent than old events. While Kraus et al. \cite{krausanalyzing} focus on users' perception of security and privacy on smartphones, it reveals psychological effects that are seen as threats from users' perspective that are usually not considered by mitigation developers.
Finally, Bauer et al.~\cite{bauer2013post} studied the relationship of time and information relevance and privacy and found that Facebook users were not really interested in the concept of ephemeral data. %

\section{Conclusion}
\label{conclusion}
With recent reports of government snooping and increasingly detrimental hacks, more and more apps have entered the market advertised as providing some privacy features. As some of these are now used by millions of users, we set to study  more carefully the features they offer. More specifically, we presented an analysis of 8 popular social networking apps namely Confide, Frankly Chat, Secret, Snapchat, Telegram, Whisper, Wickr, and Yik Yak that are marketed as offering some privacy properties. Starting from a taxonomy of 18 apps, we focused on 8 of them due to their popularity. We performed a functional, static, and dynamic analysis, aiming to analyze the properties promised by the apps. 

We found that anonymous social networks Whisper and Yik Yak actually identify their users with distinct IDs that are persistent as previous activities like chats, \textit{whispers} and \textit{yaks} are restored to the device even if the user uninstalls and reinstalls the app. This behavior shows that, although they do not require users to provide their email or phone number, they can still persistently link -- and possibly de-anonymize -- users. We also highlighted that, while Snapchat promises that messages will ``disappear'' after 10 seconds, they are not immediately deleted from its servers, as old messages are actually included in responses sent to the clients. Finally, we confirmed that apps such as Telegram and Wickr do offer end-to-end encrypted chat messaging.

In future work, we plan to extend the analysis to more  apps. We downloaded the metadata of 1.4 million apps using PlayDrone's measurements \cite{viennot2014measurement} and found 455 apps that might be offering anonymity, ephemerality, or end-to-end encryption. As it would be  demanding to manually evaluate them as we did in this paper, we will explore how to automate the analysis.

\descr{Acknowledgments.} We wish to thank Balachander Krishnamurthy for motivating our research and for numerous helpful discussions, Ruba Abu-Salma for feedback on an earlier version of the manuscript, and PRESSID for supporting Lucky Onwuzurike. This research is partly supported by a Xerox's University Affairs Committee award and the ``H2020-MSCA-ITN-2015'' Project Privacy{\&}Us project (ref.\ 675730).

\end{document}